# Dynamic photoconductive gain effect in shallow-etched AlGaAs/GaAs quantum wires

K.-D. Hof, C. Rossler, S. Manus, and J. P. Kotthaus

Department für Physik and Center for NanoScience, Ludwig-Maximilians-Universität, Geschwister-Scholl-Platz 1, D-80539 München, Germany.

A. W. Holleitner*

Walter Schottky Institut and Physik Department, Technische Universität München, Am Coulombwall 3, D-85748 Garching, Germany.

D. Schuh and W. Wegscheider

Institut für Experimentelle und Angewandte Physik, Universität Regensburg, D-93040 Regensburg, Germany.

We report on a dynamic photoconductive gain effect in quantum wires which are lithographically fabricated in an AlGaAs/GaAs quantum well via a shallow-etch technique. The effect allows resolving the one-dimensional subbands of the quantum wires as maxima in the photoresponse across the quantum wires. We interpret the results by optically induced holes in the valence band of the quantum well which shift the chemical potential of the quantum wire. The non-linear current-voltage characteristics of the quantum wires also allow detecting the photoresponse effect of excess charge carriers in the conduction band of the quantum well. The dynamics of the photoconductive gain are limited by the recombination time of both electrons and holes.

* Electronic Mail : holleitner@wsi.tum.de

Based on a proposal by Q. Hu in 1993,[1] there have been several experimental and theoretical studies on the conductive photoresponse of semiconductor quantum wires.[2],[3],[4] Part of the work focused on phenomena induced by far-infrared photons.[5] The corresponding photoresponse could be attributed to an asymmetric electron heating of the two adjacent leads of the quantum wires and thus, to a thermopower across the quantum wires.[6],[7],[8] In addition, far-infrared photons can induce a photoresponse across quantum wires that can be understood by the high-frequency rectification properties of quantum wires.[9] Recently, a persistent photoconductive gain across semiconductor quantum wires was reported for photons in the visible and near-infrared regime.[10] The persistent photoconductive gain can be understood by the capacitive influence of optically excited holes being trapped at $DX^-$ and $d^0$ centers in close vicinity of the quantum wire. Here, we report on a similar but dynamic effect which allows resolving each subband of the quantum wires as a maximum in the photoresponse. We interpret the results such that optically excited electron-hole pairs are first spatially separated because of the internal potential landscape of the quantum wires. In turn, the dynamic photoconductive gain is dominated by the optically induced holes captured at the edges of the quantum wires. There, they capacitively influence the chemical potential of the quantum wires.[11] The corresponding photoconductive gain is limited in time by the recombination time of the optically excited and spatially separated electrons and holes. We find a corresponding time constant in the millisecond regime, which is consistent with earlier findings.[12] Furthermore, the non-linear conductance characteristics of the quantum wires allow resolving the effect of the optically excited electrons as well. They induce a fast thermopower effect similar to the far-infrared case. We demonstrate the described phenomena in quantum wires which are lithographically defined in an AlGaAs/GaAs quantum well by a shallow-etch technique.[13],[14] Recently, such shallow-etched quantum wires have been exploited to demonstrate the coherent coupling of two electronic waveguide modes.[15] In principle, the presented devices are sensitive to detect single photons.[10] However, the continuous distribution of space charge in the valence band of the quantum well seems to smear out the discrete response to single photons.

Starting point is a modulation-doped AlGaAs/GaAs heterostructure which contains a quantum well comprising a two-dimensional electron gas (2DEG). The details of the heterostructure are described in [14]. As depicted in Fig. 1(a), the quantum wires are defined by a lateral constriction within the 2DEG by a combination of e-beam lithography and chemical wet etching.[13],[14],[15] Thereby, the top layer of the heterostructure is locally removed at a depth of 20 to 60 nm (see trenches A and B in Fig. 1(b)). The lithographic step locally passivates the doping layer of the heterostructure, and the 2DEG underneath is depleted (Fig. 1(c)). The lithographic width of the remaining conducting channel is approximately 350 nm. In a consecutive step, a gold gate with a thickness of 110 nm is evaporated on top of the devices. The thickness of the gate is chosen such that the gate is opaque for the utilized laser light ($\lambda_{PHOTON}$ ~ 700 – 850 nm). An aperture in the source contact defines the position, where electron-hole pairs are optically excited in the quantum well (Fig. 1(a)). The distance between the aperture and the middle of the quantum wire is approximately 4 μm. The aperture diameter of about 2 μm is larger than the optical wavelength to avoid plasmonic effects in the metal gate.[16]

All measurements are carried out in a helium continuous-flow cryostat at a vacuum of about $10^{-5}$ mbar and a bath temperature of 2.0 K. For the photoresponse measurements, the charge carriers are locally excited by focusing the light of a mode-locked titanium:sapphire laser with a repetition rate of 76 MHz onto the aperture of the samples. The power density at the position of the aperture is chosen to be ~ 300 mW/cm$^2$ (at a photon energy $E_{PHOTON}$ = 1.55 eV). The two-terminal differential conductance across the quantum wires is determined by a standard lock-in technique. Applying a voltage to the gate, the chemical potential of the subbands in the quantum wires is shifted with respect to the Fermi energies of the source/drain contacts. Typical for spin degenerate quantum wires, the conductance $G^{OFF}$ of the devices with the laser being "off" shows quantized steps of 2e$^2$/h for each subband (black squares for sample A in Fig. 1(d)).[2],[3] Illuminating the aperture of the devices with laser light, we find that the pinch-off voltage of the conductance $G^{ON}$ of the devices with the laser being "on" is shifted to a more negative bias by a value $\Delta V_G$ depending on the laser power (open symbols in Fig. 1(d)). The sign of the shift indicates that optically excited holes capacitively influence the chemical potential of the quantum wire. To resolve the dynamics of the photoresponse of the quantum wires, we chop the laser at frequency $f_{CHOP}$. The resulting ac-photoresponse $|I_{PR}| = |I_{PR}(E_{PHOTON}, f_{chop})| = |I^{ON}(E_{PHOTON}, f_{chop}) - I^{OFF}(E_{PHOTON}, f_{chop})|$ across the sample with the laser being in the "on" or "off", respectively, is amplified by a current-voltage converter and detected with a lock-in amplifier utilizing the reference signal provided by the chopper. Fig. 1(d) depicts such a photoresponse measurement of sample A as a function of $V_G$, while the aperture of sample A is illuminated. We find that the photoresponse shows a maximum at the onset of each one-dimensional subband of the quantum wire.

Generally, sample A (sample B) exhibits conductance characteristics which are typical for lithographically defined quantum wires with a subband energy of about 4-6 meV (2 meV).[13],[15] Fig. 2(a) shows the conductance $G^{OFF}$ of sample A as a function of the source/drain voltage $V_{SD}$ for certain gate voltages $V_G$, while the laser is being "off". The corresponding transconductance $dG^{OFF}/dV_G$ is depicted in Fig. 2(b). In both graphs, one can clearly identify the integer and half-integer plateaus, which are related to the alignment of the one-dimensional subbands of the quantum wires to the chemical potentials of the source/drain contacts.[13],[15] Intriguingly, the corresponding diamond structure is also revealed in the photoresponse measurements (grayscale plot for sample A in Fig. 2(c)). Fig. 2(d) illustrates the photoresponse of sample B as a function of $V_G$ and $V_{SD}$. Again, we find a diamond structure for each subband of the quantum wire. The photoresponse of sample A as a function of $V_{SD}$ is depicted in Fig. 2(e) (the trace corresponds to the dotted line in Fig. 2(c)). We would like to note following points. First, the offset of $V_{SD} \sim -40$ μV at $|I_{PR}| \approx 0$ is caused by an input-offset of the current-voltage converter. Since it depends neither on the laser power nor on the photon energy, we can exclude rectification effects to cause the photoresponse.[9] Second, for $|V_{SD}| \leq 2$ mV the photoresponse is directly proportional to $V_{SD}$. Both findings are consistent with the interpretation of a dominating photoconductive gain effect, which is induced by the capacitive influence of optically excited holes stored in close vicinity of the quantum wires.[10],[11] In the present case, the non-linear I-V characteristics of the quantum wires of samples A and B allow exploring the photo-induced effects further. To this end, we highlight the photoresponse of sample B as a function of $V_G$ for different $V_{SD}$ in Fig. 2(f). We find that the photoresponse maxima can be resolved also for the smallest value of $V_{SD}$. We interpret the finding of a finite photoresponse at zero bias as the fingerprint of an electronic thermopower effect, which will be discussed below.

Fig. 3(a) depicts the spectral dependence of the photoresponse maxima of sample A. The quantum wire shows photoresponse for $E_{PHOTON} > 1.50$ eV with a spectral maximum at $E_{PHOTON} \sim 1.55$ eV. Latter corresponds to the band gap energy of the quantum well, as can be seen by the photoluminescence peak in Fig. 3(a). Hereby, we can assume that the photoresponse is caused by charge carriers in the quantum well and not in the adjacent GaAs or AlGaAs layers of the heterostructure.[17] The relatively broad width of the photoresponse maximum is consistent with the spectral width of the utilized laser of about 11 meV. Fig. 3(b) shows the dependence of the photoresponse on the chopper frequency. By fitting exponential decay curves to the experimental data (Fig. 3(b)), we find two time constants of $t_1 \sim 11.2 \pm 1.6$ ms and $t_2 \sim 2.4 \pm 0.1$ ms for the photoresponse across the quantum wire. In principle, both time constants are consistent with the recombination time of optically induced and spatially separated holes and electrons in laterally structured semiconductor devices.[12] In order to interpret both time constants further, we introduce the following simple transport model.

Optically created electrons and holes can contribute differently to the photoresponse of semiconductor wires. In the present case, due to the saddle-point potential of quantum wires,[13] the holes in the valence band of the quantum well are stored at the edge of the constriction spatially separated from the electron channel (see

Fig. 1(a)).[11] There, they act as a local positive space charge.[10] In turn, the conductance trace of the quantum wires is shifted by an additional $\Delta V_G$ with respect to the gate voltage without the presence of the holes (Fig. 1(d)). In other words, the subbands of the quantum wires are shifted in energy by a value of $\Delta E = \alpha \cdot e \cdot \Delta V_G$ (1) at the presence of the holes. The average capacitive coupling coefficient $\alpha$ can be experimentally determined from the slopes of the diamond structures in Fig. 2(b) to be ~ 0.2.[18] At the same time, the optically induced electrons can increase the electron temperature $T_S$ and sheet density $n_s$ in the source region of the devices (Fig. 1(a)). The local temperature increase in the source contact gives rise to a thermopower across the quantum wire, since the electron temperature $T_D$ in the drain contact can be assumed to be constant.[6]-[8]

Combining both mechanisms, Fig. 4(a) depicts a sketch of our simple model to describe the experimental situation with $t = t(E, V_G, \Delta V_G)$ the Heavyside step function as the transmission function of the quantum wire and $\mu_S$ and $\mu_D$ the chemical potentials of the source and drain electrodes. To describe the effect of both holes and electrons, we calculate the conductance of the quantum wire as follows (2):

$$G(V_G) = \frac{1}{V_{SD}} \int_{-\infty}^{+\infty} t(E, V_G, \Delta V_G) \cdot [f(E, \mu_S, T_S) \cdot (1 - f(E, \mu_D, T_D)) - f(E, \mu_D, T_D) \cdot (1 - f(E, \mu_S, T_S))] dE .$$

For the conductance traces $G^{ON}$ and $G^{OFF}$ in Fig. 4(b), we assume following parameters for the laser being "on" or "off": $T_S^{ON}$ = 9.5 K, $T_S^{OFF}$ = 6.5 K, $T_D^{ON} = T_D^{OFF}$ = 6.5 K, $V_{SD}$ = 1 mV, and $\Delta V_G$ = 25 mV. We find that our model can reproduce the data of sample A reasonably well (Fig. 4(c)). In Fig. 4(d) and (e) we compare the calculated difference $\Delta G = G^{ON} - G^{OFF}$ as in Fig. 4(b) and (c). Again, this static case is reproduced by our model reasonably well. We would like to note that the asymmetric form of the peaks in Fig. 4(e) is represented in our model by unequal electron temperatures $T_S$ and $T_D$ in equation (2). This fingerprint of a thermopower effect, which is induced by the optically excited electrons, is consistent with the fact that we resolve photoresponse peaks also at the smallest values of $V_{SD}$ (Fig. 2(f)).

In order to model the ac-photoresponse across the quantum wires, we introduce following expression (3): $\Delta G = G^{ON} - G^{OFF}_{INTERMEDIATE}$, with $G^{OFF}_{INTERMEDIATE}(V_G) = G^{OFF}(V_G + \delta \Delta V_G)$ the intermediate conductance state of the quantum wire for the laser being "off" in the time interval $t_{CHOP} = 1/f_{CHOP}$. Fig. 4(f) presents model calculations for $\delta = 0.5$, 0.9, and 0.99 in equation (3). In other words, within the given time interval $t_{CHOP}$ the intermediate state is shifted back in gate voltage by a fraction 50%, 10%, and 1% as compared to the shift $\Delta V_G$ between $G^{ON}$ and the equilibrated off-state $G^{OFF}$. The model describes the ac-photoresponse reasonably well for $f_{CHOP} < 500$ Hz. In this frequency regime, we experimentally observe that the maxima of the photoresponse peak shift towards more positive gate voltages when decreasing the chopper frequency (see open circles in Fig. 4(g)). We interpret the finding such that for a longer time interval $t_{CHOP}$ less holes are still captured at the edges of the quantum wires. Hereby, the intermediate conductance "off"-state $G^{OFF}_{INTERMEDIATE}$ is shifted to more positive gate voltages for longer $t_{CHOP}$. In other words, we interpret the first time constant $t_1 \sim 11.2 \pm 1.6$ ms (Fig. 3(b)) to be the recombination time of the optically induced and spatially separated electrons and holes. For $f_{CHOP} > 500$ Hz, we experimentally find the opposite: the maxima of the photoresponse peak shift towards more positive gate voltages for increasing the chopper frequency (data not shown). We interpret the second shift such that the conductive state $G^{ON}$ of the quantum wire is not built up entirely during the time interval $t_{CHOP}$. Therefore, the second time constant $t_2 \sim 2.4 \pm 0.1$ ms (Fig. 3(b)) describes the build-up dynamics of the space charge, which gives rise to the photoconductive gain effect. We experimentally observe that both time constants $t_1$, and $t_2$ slightly depend on the laser power (data not shown); reflecting the complex charge dynamics at the edges of the quantum wires. Finally, we would like to note that equations (1), (2), and (3) also reproduce the linear dependence of the photoresponse on $V_{SD}$ (Fig. 2(e)).

In summary, we present photoresponse measurements on semiconductor quantum wire devices. We interpret the results by a dynamic photoconductive gain effect, i.e. optically induced and spatially separated charge carriers capacitively influence the chemical potential of the one-dimensional subbands of the quantum wire. By varying the chopper frequency of the exciting laser, we find two time constants of the photoresponse across the quantum wire. We interpret the first time constant of about $\sim 11$ ms to be the recombination time of space-separated electron and holes, while the second time constant of about $\sim 2$ ms describes the interplay between the chopper frequency of the laser and the build-up dynamics of the space charge giving rise to the photoconductive gain effect. The experimental photoresponse is reproduced by a simple transport model. The model is consistent with the interpretation that the photoresponse across the quantum wires is dominated by a photoconductive gain effect induced by optically excited holes located in close vicinity of the quantum wires. At the same time, the model allows distinguishing the photoconductive gain effect from a thermopower effect, which is induced by the optically excited electrons. In principle, the described devices are sensitive to detect single-photons.[10] However, the continuous distribution of space charge in the valence band of the quantum well in the vicinity of the quantum point contact smears out the discrete response to single charges and photons.

We gratefully acknowledge financial support from BMBF via nanoQUIT, the DFG (Ho 3324/4), the Center for NanoScience (CeNS), the LMUexcellence program and the German excellence initiative via the "Nanosystems Initiative Munich (NIM)". In addition, we thank J. Simon for technical support.

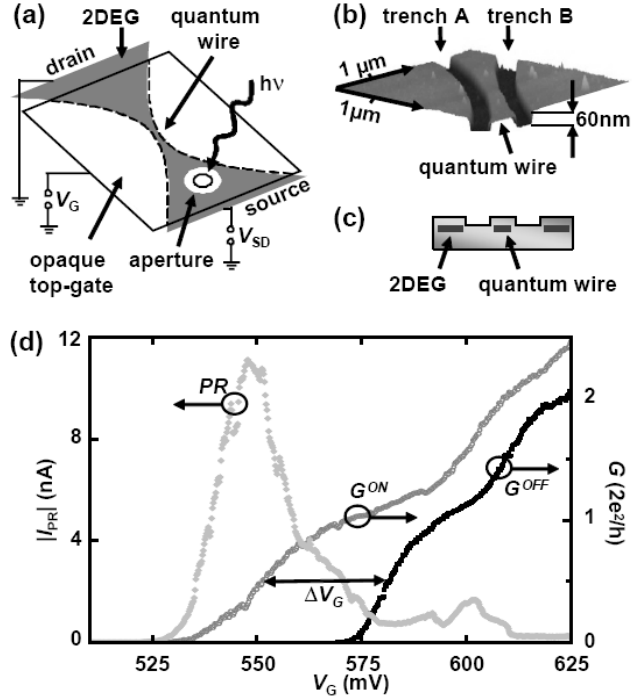

Fig. 1(a): Experimental circuit. A lateral constriction of a two-dimensional electron gas (2DEG) forms a quantum wire between source and drain contacts. The central area of the device is covered with an opaque gate (bright rectangle). An aperture in the gate close to the constriction defines the position where the underlying 2DEG is optically excited. (b) and (c): The quantum wire is defined by two adjacent trenches A and B in the 2DEG via a shallow- etch technique. (d) Conductance and photoresponse data of sample A at $T_{BATH}$ = 2.3 K, $f_{CHOP}$ = 117 Hz and $V_{SD}$ = 1.5 mV. See text for details.

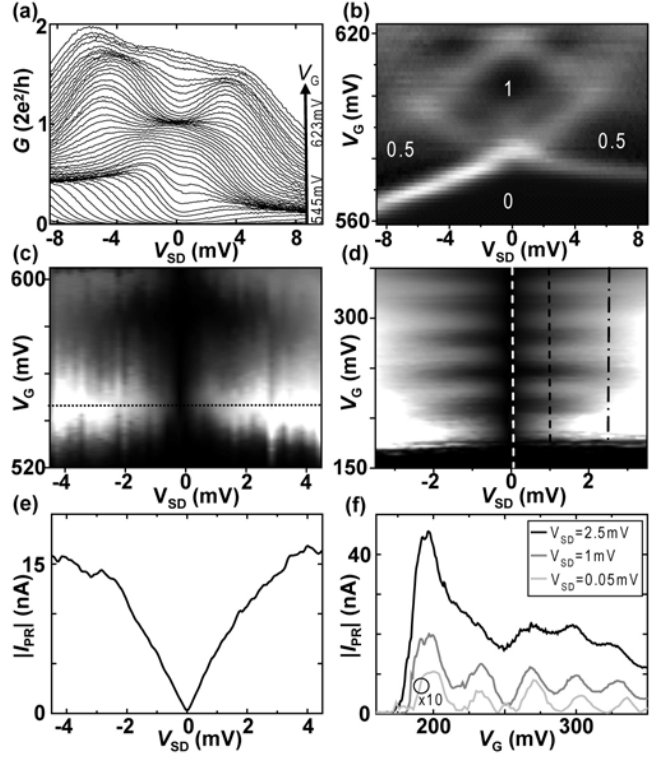

Fig. 2(a): Conductance data of sample A as a function of $V_{SD}$ for gate voltages in the range of $V_G$ = 545 mV and 623 mV ($T_{BATH}$ = 2.1 K). (b) Corresponding linear gray-scale plot of the absolute transconductance data (black = zero transconductance, white = high absolute value) (c) Linear grayscale plot of the photoresponse of sample A as a function of $V_{SD}$ and $V_G$ at $T_{BATH}$ = 2.3 K (black = 0 nA, white = 25 nA). (d) Linear grayscale plot of the photoresponse of sample A as a function of $V_{SD}$ and $V_G$ at $T_{BATH}$ = 2.2 K (black = 0 nA, white = 55 nA). (e) Single photoresponse curve which corresponds to the dotted line in Fig. 2(c). (f) Photoresponse curves which correspond to the white-dashed, black-dashed, dashed-dotted lines in Fig. 2(d). See text for details.

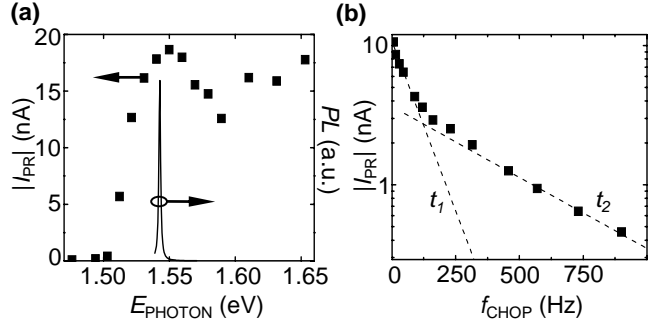

Fig. 3(a): Squares: photoresponse $|I_{PR}|$ of sample A as a function of excitation photon energy $E_{PHOTON}$ with a maximum at $E_{PR} \sim 1.55$ eV. Line graph: spectrally resolved photoluminescence $PL$ of the quantum well with a maximum $E_{PL} = 1.545$ eV. Data taken in a microPL setup at $T_{2DEG} \sim 10$ K. (b) Photoresponse $|I_{PR}|$ of sample A at the first photoresponse maximum in Fig. 1(d) as a function of the chopping frequency $f_{CHOP}$ in a logarithmic representation. The two lines are fits to exponential decay functions. See text for details. $T_{BATH} = 2.7$ K, $P = 80$ mW/cm$^2$.

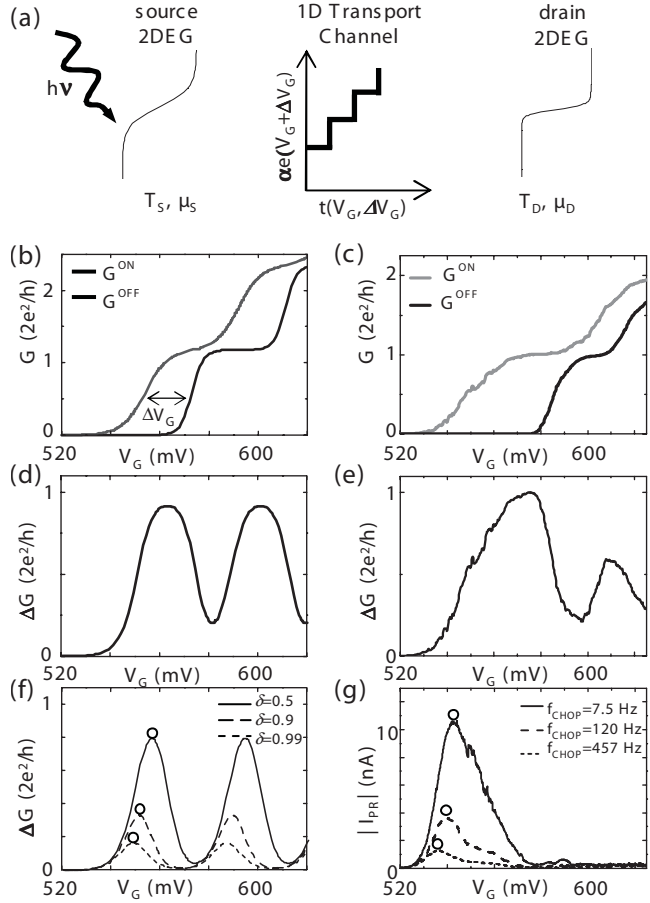

Fig. 4(a): Simple model to describe the photoresponse of the quantum wire: the 2DEGs in the source and drain contacts are described by Fermi-distributions, while the transfer function of the quantum wire is described by a heavyside step-function. (b), (d), and (f): Calculated results of equation (1), (2) and (3). (c) and (e): Experimental data at $T_{BATH}$ = 2.0 K  (g) Experimental data at $T_{BATH}$ = 2.7 K. See text for details.